\documentclass[11pt,a4paper]{article}
\usepackage{jheppub_kim}
\usepackage{epsfig}
\usepackage{amsmath}
\usepackage{graphicx}
 \usepackage{graphics}
 \usepackage[scriptsize,nooneline,hang]{caption}
\usepackage[hang,nooneline,scriptsize]{subfigure}
 
\begin{document}

\title{Thermodynamic geometry of a black hole surrounded by perfect fluid in Rastall theory}

\author[a]{Saheb Soroushfar, }
\affiliation[a]{Faculty of Technology and Mining, Yasouj University, Choram 75761-59836, Iran}
\author[b]{Reza Saffari, }
\affiliation[a]{Department of Physics, University of Guilan, 41335-1914, Rasht, Iran}
\author[c,d]{Sudhaker Upadhyay }
\affiliation[c]{Department of Physics, K.L.S. College, Nawada-805110, India}
\affiliation[d]{Visiting Associate, Inter-University Centre for Astronomy and Astrophysics (IUCAA) Pune, Maharashtra-411007}
\emailAdd{soroush@yu.ac.ir}
\emailAdd{rsk@guilan.ac.ir}
\emailAdd{sudhakerupadhyay@gmail.com}

\abstract{In this paper, we study thermodynamics and thermodynamic geometry of a 
black hole surrounded by the perfect fluid in Rastall theory. In particular, we calculate the physical
quantity like mass, temperature and heat capacity of the system for two different cases.  
From the resulting heat capacity, we emphasize stability of the system.  
Following Weinhold, Ruppiner, Quevedo and HPEM formalism, thermodynamic geometry of  this black hole 
in Rastall gravity is  also analyzed. We find that the singular points of the curvature scalar of
Ruppeiner and HPEM metrics entirely coincides with zero points of the heat capacity.
But there is another divergence of HPEM metric which coincides with the singular points of heat capacity, 
so we can extract more information of HPEM metric compared with Ruppeiner metric.
However, we are unable to find any physical data about the system from the Weinhold and Quevedo formalism.
}

\maketitle

\section{Introduction}  

In understanding theory of gravitation, general relativity (GR) follows the covariant conservation of matter energy-momentum tensor. The limitation of  such idea is that   conservation of  energy-momentum tensor has been probed only in the flat or weak-field arena of spacetime. A generalization of this theory, so-called Rastall theory, has recently been proposed, which relaxes the necessity of
  covariant conservation of the energy-momentum tensor by adding  some new terms to the Einstein's equation \cite{ras}.  Recently, this idea which has been supported  in Ref. \cite{jos}, which shows that the divergence of the energy-momentum tensor can be non-zero in a
curved spacetime. Rastall theory provides an explanation for inflation problem and other cosmological problems \cite{3,4,5,6,7,8}.  
In order to explain accelerated expansion of universe, one may think that the negative pressure is generated by some peculiar kind
of perfect fluid, where the proportion between the pressure and energy density is between
$−1$ and $−1/3$. If the scalar field generates this perfect fluid, it is generally called
quintessence.  In Rastall theory of gravitation, the various  black hole solutions surrounded by perfect fluid has been discussed \cite{9,10,11,12,13,Soroushfar:2018yzw}.

Hawking and Bekenstein were  first who  proposed that  black holes have thermodynamic 
properties \cite{haw,bak}. There after, this subject extensively studied by many people  \cite{13,14,15,16,17,18,19,Soroushfar:2016nbu,Rodrigue:2018lzp}. 
In this connection, it has been found that the thermodynamic properties of black holes in Rastall gravity  investigate 
the nature of a non-minimal coupling between geometry and matter fields. 
Recently, thermodynamics of various black holes under the effect of small thermal 
fluctuations have also been studied extensively.  For example, thermodynamics of Van der 
Waals black holes \cite{20}, static black hole in f(R) gravity \cite{Upadhyay:2018bqy}, charged rotating AdS black holes \cite{21}, 
black holes in gravity’s rainbow \cite{22,23},  quasitopological black holes \cite{24},
 massive gravity black hole 
 \cite{sud},   Ho\v rava-Lifshitz black hole 
 \cite{sud1} and  Schwarzschild-Beltrami-de Sitter black hole   \cite{sud2} have been discussed.

Meanwhile,  several attempts have been made to implement differential geometric concepts in thermodynamics of black holes. 
In an attempt, to formulate the concept of thermodynamic
length, Ruppeiner  metric \cite{Ruppeiner}, which conformally equivalent to Weinhold's
metric \cite{Weinhold}, is introduced. Furthermore, the phase space and the metric structures suggested that Weinhold's and Ruppeiner's  metrics are not invariant under Legendre transformations \cite{sal,mru}. Afterwards, another formalism is developed by Quevedo \cite{Quevedo},  
which unifies the geometric properties of the phase space and the
space of equilibrium states.  
Also, a new metric (HPEM metric) \cite{Hendi:2015rja,Hendi:2015xya,EslamPanah:2018ums} 
was introduced in order to build a geometrical phase space by thermodynamical quantities. 
Though, several study of thermodynamics of various black holes exploiting thermodynamic geometric methods have been made, the  thermodynamics of black holes  surrounded by perfect fluid in Rastall theory  remains unstudied yet.  This provides us an opportunity to bridge this gap. This is the motivation of present study. 
 
In this paper, we consider a black hole solution surrounded
by the (general)  perfect fluid in Rastall theory and describe its  thermodynamics by means of two specific cases. In this regard, we first compute mass  by setting the metric function representing black hole surrounded
by the quintessence field to zero. Furthermore, by exploiting standard thermodynamic  relations,
we derive the Hawking temperature and heat capacity.  The heat capacity  plays a pivotal role in  describing stability of the black hole. In order to study the behavior of the resulting thermodynamic entities, we plot graphs for them in terms of horizon radius.
By doing so, we find that  
the temperature of black hole  surrounded
by  the   quintessence field in Rastall theory remains positive only in a
particular range of event horizon for quintessence factor $ N_q = 0.05$.  The larger values of quintessence factor decrease temperature of the system. This suggests that increasing in the values of quintessence the physical area of this black hole decrease. The heat capacity plot shows that 
there are physical limitation points and phase transition critical points occurring at different values of event horizons. For this black hole, stability occurs for some particular values horizon radius. We also study the
geometric structure of the black hole surrounded by quintessence field by calculating the Weinhold, Ruppiner, Quevedo and HPEM curvature scalars. We find that Weinhold curvature scalar  vanishes and becomes fail  to describe the
phase transition. We obtain a non-zero curvature scalar for the Ruppiner and HPEM geometries and, in these 
cases, the singular points coincide with zero points of the heat capacity. Also for the HPEM formalism, 
there is another divergency of HPEM metric which coincides with the singular points of heat capacity and therefore we can extract more information  rather than other mentioned metrics.
The curvature 
scalar of Quevedo metric is also calculated. Here, in this case we can't find any physical data about the system.

Moreover, thermodynamic properties of black hole surrounded by dust field in Rastall 
gravity are also studied. The mass, Hawking temperature and heat capacity of
black hole in dust field background are calculated. In order to do comparative analysis, we
plot graphs in terms of horizon radius. From the obtained plots, it is obvious  that  the mass of the system has one minimum point at  particular value of horizon radius. The behavior of  temperature with horizon radius is similar to  the quintessence case.
For instance, increasing the values of dust factor ($N_d$) the physical area of this black hole decrease. 
The heat capacity plot tells that stability of black hole increases with the size of black hole upto a certain point. After that point a phase transition occurs
and makes the black hole more unstable. We calculate the 
curvature scalar to discuss the thermodynamic geometry for this case also. 

The paper is presented in the following manner. 
In section \ref{section3}, we revisit the basic setup of black hole surrounded by perfect fluid in Rastall theory. Thermodynamics and thermodynamic geometry of black hole
surrounded by quintessence field are discussed in section \ref{section4}. Thermodynamics and thermodynamic geometry of black hole
surrounded by  dust field is presented in section \ref{section5}. We summarize results and discussions in the last section. 
 
\section{Field equations of a black hole surrounded by perfect fluid in Rastall theory}\label{section3}
In this section, we briefly review field equations and metric components in the context of Rastall theory. 
The Rastall field equations for a space-time with Ricci scalar $ R $ and an energy momentum source of
$ T_{\mu \nu}$ can be written as
\begin{equation}
G_{\mu \nu}+k \lambda g_{\mu \nu}=k T_{\mu \nu},
\end{equation}
where $ k $ and $ \lambda $ are the Rastall gravitation coupling constant and the Rastall parameter 
which indicates deviation scale from the standard GR, respectively. 
The metric of a black hole surrounded by perfect fluid in Rastall theory would be deduced as follows \cite{12}
\begin{equation}\label{mainmetric}
ds^2=-f_{s}(r)dt^2+\frac{dr^2}{f_{s}(r)}+r^2(d\theta^2+\sin^2\theta d\phi^2),
\end{equation}
with
\begin{equation}\label{eqfr}
f_{s}(r)=1-\frac{2M}{r}+\frac{Q^2}{r^2}-\frac{N_s}{r^{\frac{1+3w_s-6k\lambda(1+w_s)}{1-3k\lambda(1+w_s)}}}.
\end{equation}
where  subscript ``s" represents the surrounding field, $N_s$ is surrounding field structure parameter, $Q$ is charge, 
$w_s$ is equation of state parameter and $M$ is mass of the black hole.  
By considering $ \lambda=0 $ and $ k=8\pi G_{N} $, this metric regains the Reissner-Nordstr\"om black hole 
surrounded by a surrounding field as \cite{kiselev} 
\begin{equation}\label{metrick}
ds^2=-\left(1-\frac{2M}{r}+\frac{Q^2}{r^2}-\frac{N_s}{r^{3w_s+1}}\right)dt^2+\frac{dr^2}{\left(1-\frac{2M}{r}+\frac{Q^2}{r^2}-\frac{N_s}{r^{3w_s+1}}\right)}+r^2d \Omega ^2.
\end{equation}
We use the metric (\ref{mainmetric}) to investigate the thermodynamic geometry of a black hole surrounded by perfect fluid in Rastall theory.
In particular, we consider the quintessence and dust fields as sub-classes of the Rastall fields.
\section{Black hole surrounded by the quintessence field}\label{section4}
In this section, we are going to investigate the thermodynamic properties of a black
hole surrounded by the quintessence field. A quintessence field plays important role for the observed accelerated expansion of the universe \cite{kiselev}.  By considering $w_s=w_q=-\frac{2}{3}$, the metric (\ref{mainmetric}) takes the following form:
\begin{eqnarray}\label{quin}
&ds^2=-f_q(r)dt^2+\frac{dr^2}{f_q(r)}+r^2d\Omega ^2,\\
&f_q(r)=1-\frac{2M}{r}+\frac{Q^2}{r^2}-\frac{N_q}{r^{\frac{-1-2k\lambda}{1-k\lambda}}}.
\end{eqnarray}
For the case $k\lambda=\frac{1}{4}$, metric (\ref{quin}) can be written as
\begin{eqnarray}\label{RQfr}
&ds^2=-f_q(r)dt^2+\frac{dr^2}{f_q(r)}+r^2d \Omega ^2,\\
&f_q(r)=1-\frac{2M}{r}+\frac{Q^2}{r^2}-N_qr^2.
\end{eqnarray}
\subsection{Thermodynamics} 
\begin{figure}[h]
	\centering
	 \subfigure[$ N_{q}=0.05 $]{
		\includegraphics[width=0.38\textwidth]{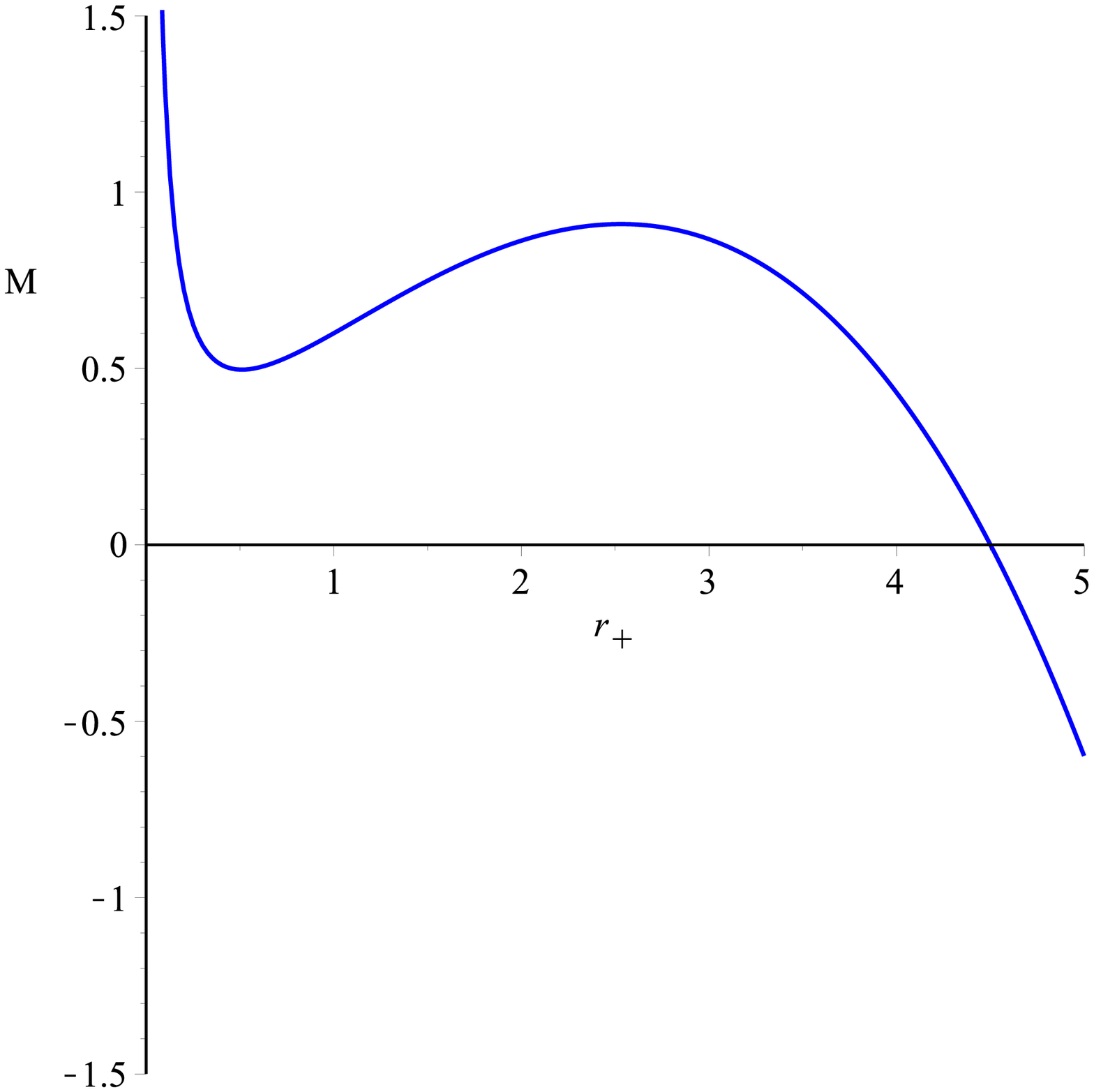}
	}
      \subfigure[$ N_{q}=0.05 $]{
		\includegraphics[width=0.38\textwidth]{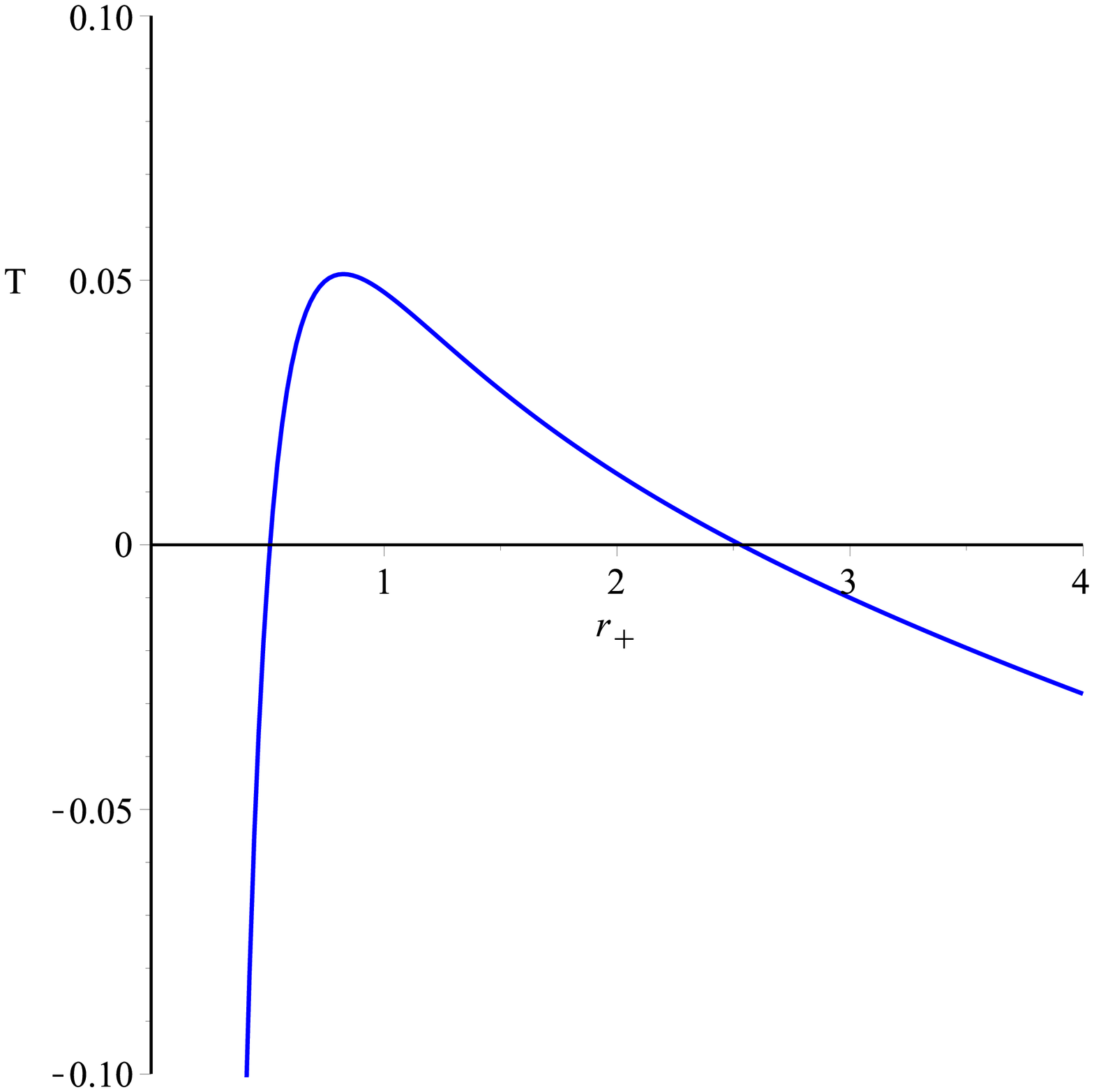}
	}
     \subfigure[]{
		\includegraphics[width=0.38\textwidth]{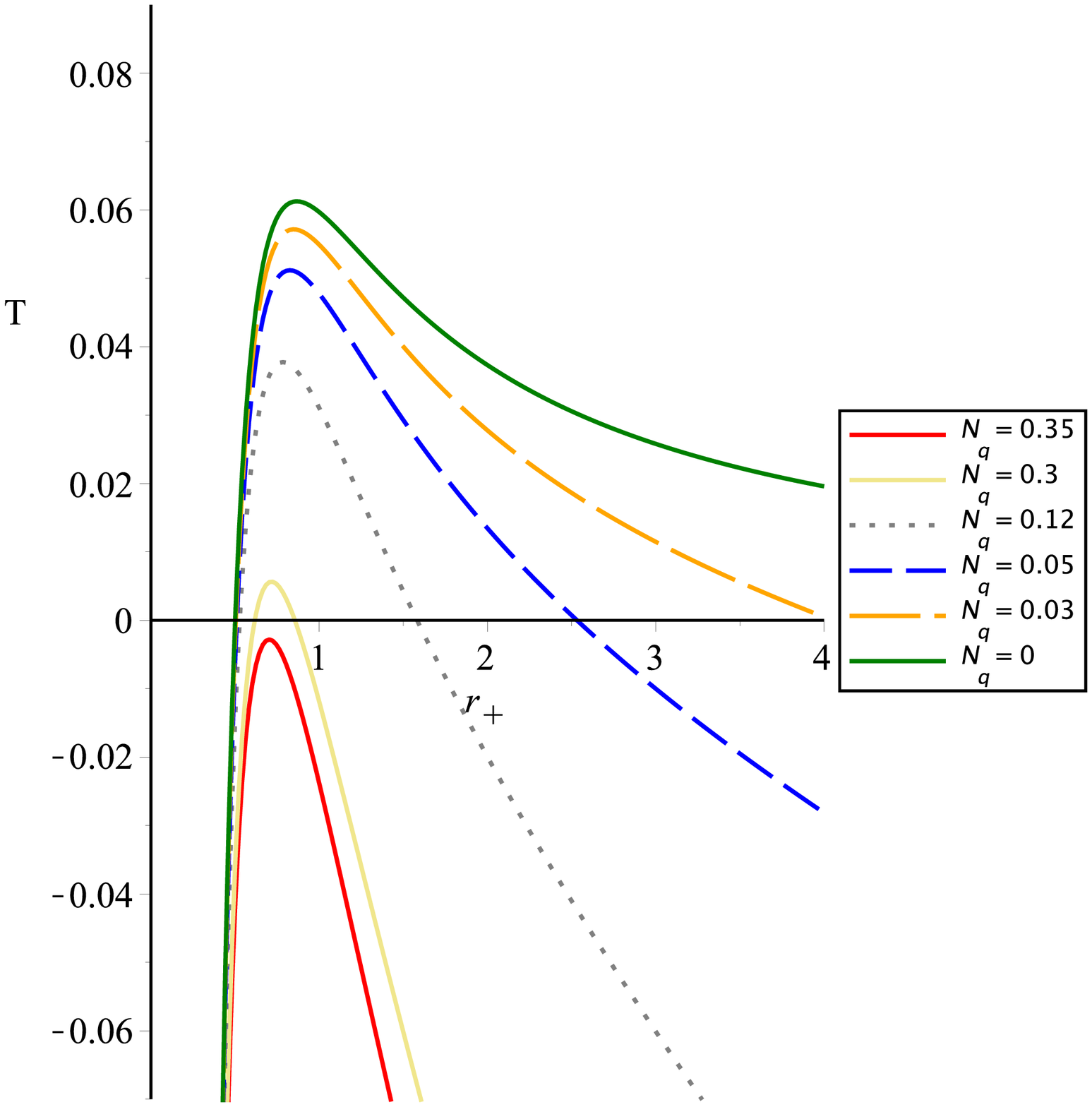}
	}
    \subfigure[]{
		\includegraphics[width=0.38\textwidth]{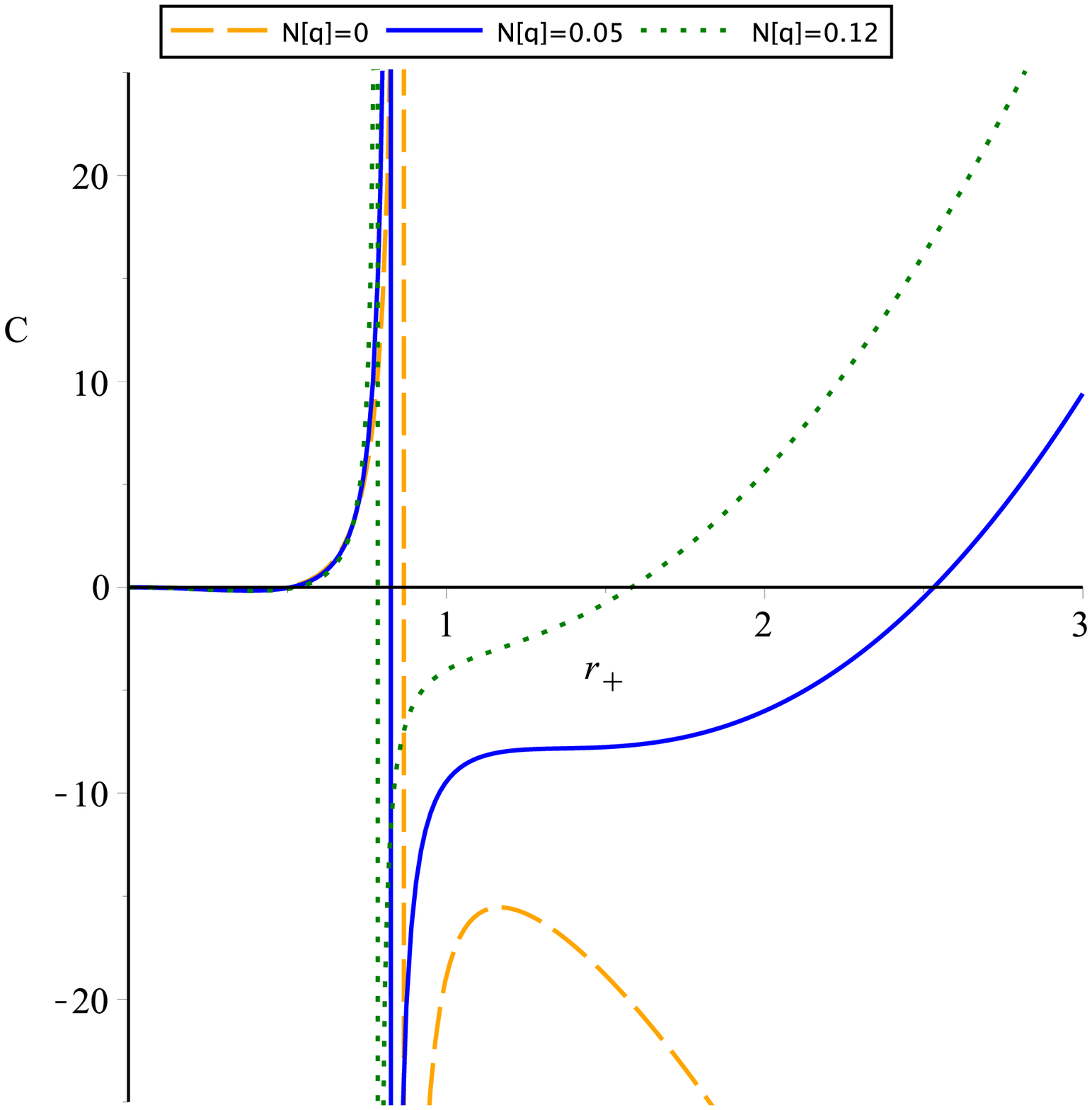}
	}
     \subfigure[$ N_{q}=0.05 $]{
		\includegraphics[width=0.38\textwidth]{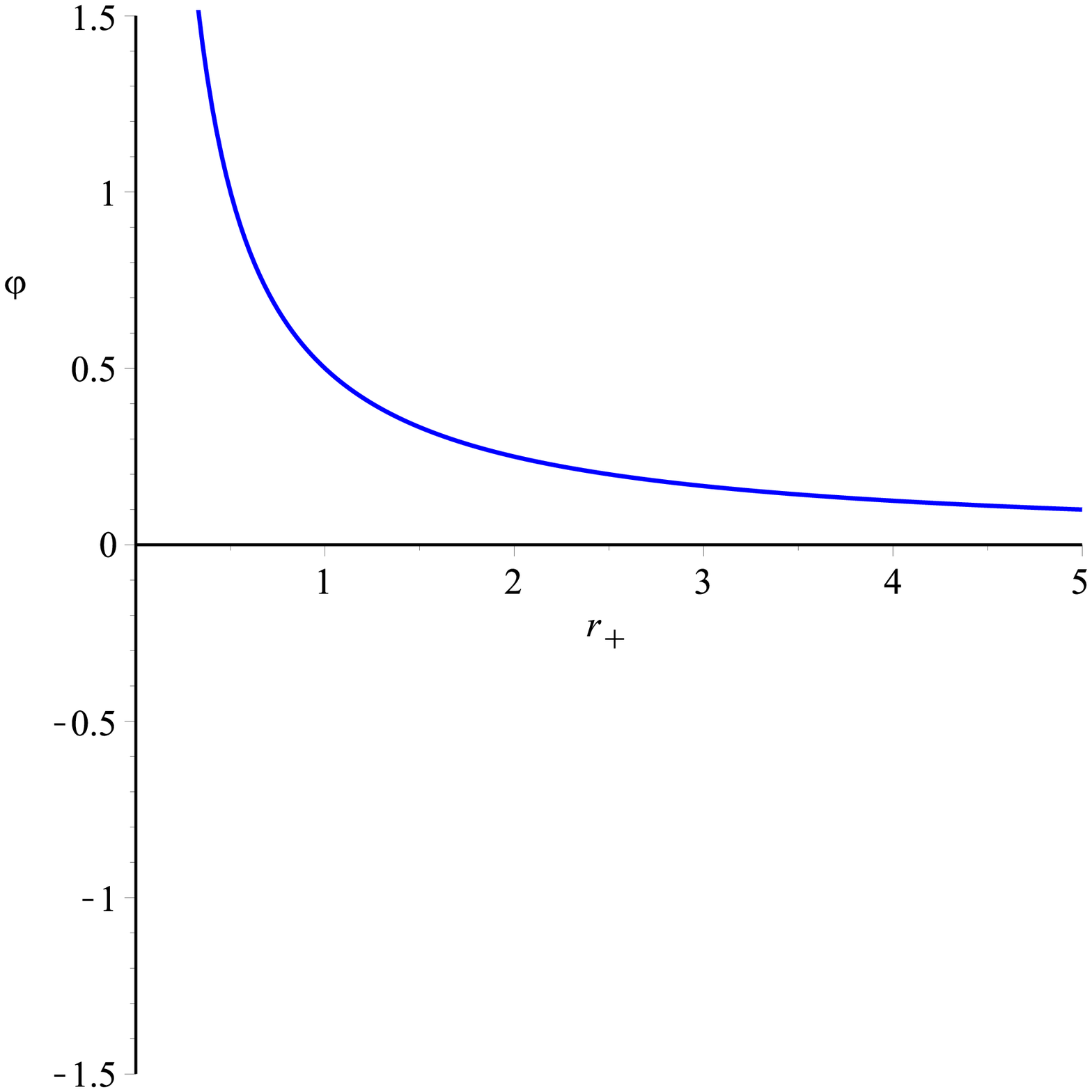}
	}
	\subfigure[$ N_{q}=0.05 $]{
		\includegraphics[width=0.38\textwidth]{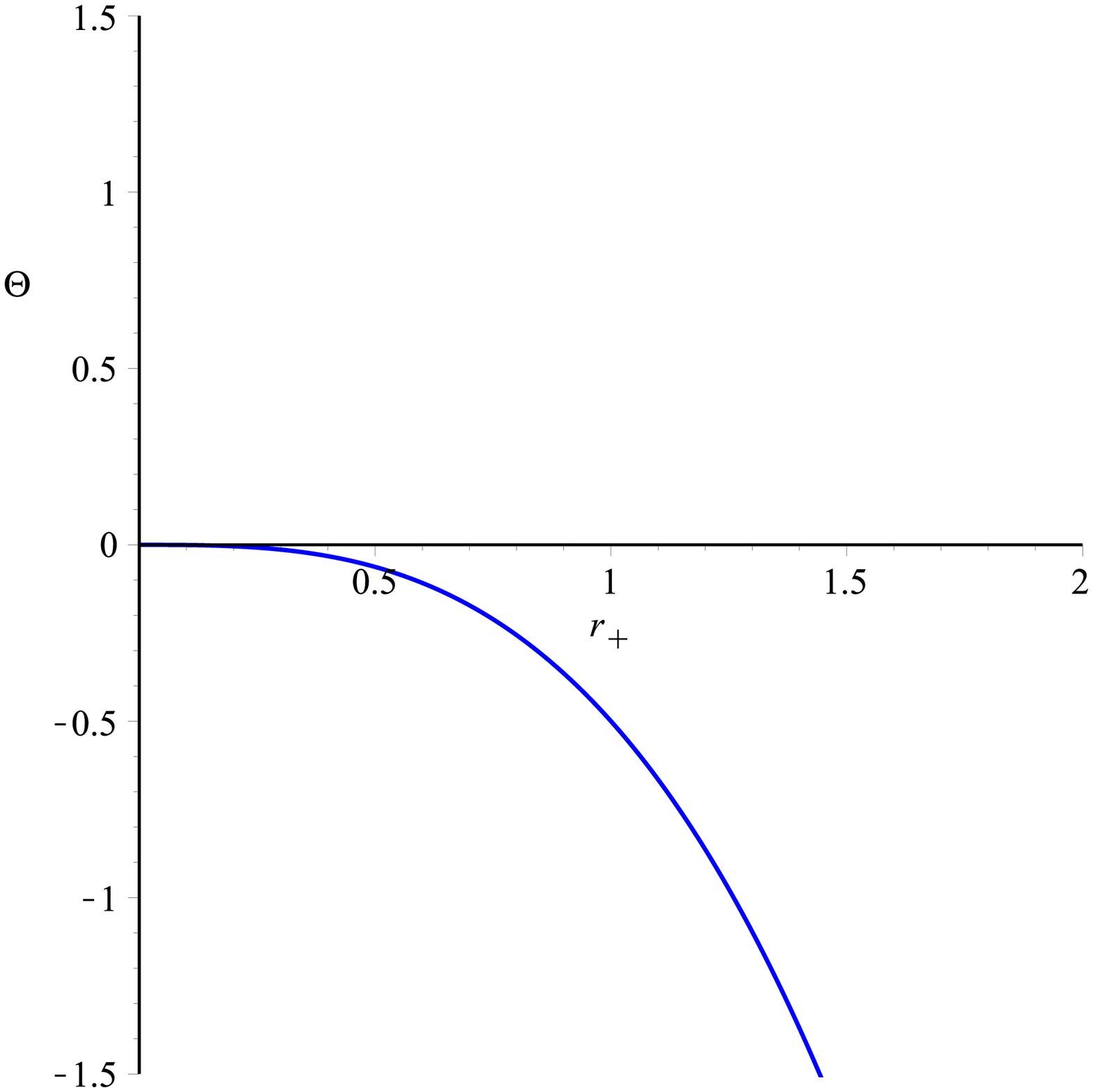}
	}
	\caption{Variations of thermodynamic parameters of a black hole surrounded by 
	 quintessence field in terms of horizon radius $ r_{+} $ for $ Q=\sqrt{0.25} $.}
 \label{pic:RQParameters}
\end{figure}

The mass corresponding to Eq. (\ref{RQfr}), can be obtain using $ f_q({r})=0$.
Also, we can express the mass of the black hole $M$, in terms of its
entropy $S$, using the relation between entropy $S$  and event horizon radius
$r_{+}$  $ (S=\pi r^{2}_{+})$, as 
\begin{equation}\label{RQmass}
M(S,N_{q},Q)=\frac{{{\pi ^2}{Q^2} + \pi S - N_{q} {S^2}}}{{2{\pi ^\frac{3}{2}}S^{\frac{1}{2}} }} .
\end{equation}
First law of thermodynamics for a black
hole surrounded by quintessence field can be written as 
\begin{equation}\label{RQlow}
{\it dM}={\it TdS}+\Psi \,dN_{q}+\varphi {\it dQ} ,
\end{equation}
where $ \varphi $ is a quantity conjugate to electric charge $ Q $ and 
$ \Psi $ is a quantity conjugate to quintessence parameter $ N_{q} $. 
So, by using the Eqs. (\ref{RQmass}) and (\ref{RQlow}), we   obtain thermodynamic parameters like
  temperature $(T=\frac{\partial M}{\partial S})$, heat capacity $(C=T\frac{\partial S}{\partial T})$, 
$(\Psi=\frac{\partial M}{\partial N_{q}})$, and $(\varphi=\frac{\partial M}{\partial Q})$ as follows
\begin{equation}
T=- \frac{{{\pi ^2}{Q^2} - \pi S + 3 N_{q} {S^2}}}{{4{\pi ^\frac{3}{2}}{S^\frac{3}{2}}}},
\end{equation}

\begin{equation}
C= - \frac{{2 S \left( {{\pi ^2}{Q^2} - \pi S + 3 N_{q} {S^2}} \right)}}{{3{\pi ^2}{Q^2} - \pi S - 3 N_{q} {S^2}}},
\end{equation}

\begin{equation}
\varphi=\sqrt {\frac{\pi}{S}}Q  ,
\end{equation}

\begin{equation}
\Psi=-\frac{1}{2}\,\left(\frac{S}{\pi}\right)^{3/2} .
\end{equation}
These thermodynamic parameters are plotted in terms of horizon radius $r_+$, (see Fig.~\ref{pic:RQParameters}).
\\
From  Fig.~\ref{pic:RQParameters}(a), we find that for $N_{q}=0.05$, the mass of this black hole has
a minimum value at $ r_{1}= 0.51$, afterword it arrives to its maximum value at $ r_{2}=2.53$, 
then it becomes zero at $ r_{0}=4.5$. 
Also, the behaviour of the temperature versus  $ r_+ $  is demonstrated in Fig.~\ref{pic:RQParameters}(b). 
 From this figure, we see that  for low values of $ r_+ $, 
the temperature increases to a maximum point and then it starts decreasing with higher $ r_+$. 
In  other words, temperature is positive only
in a particular range of event horizon $ (r_{1}<r_+<r_{2}) $ for $N_{q}=0.05$, however, at $ r<r_{1} $, and $ r_{2}<r $, 
it will be negative and leads to an unphysical solution. 
In addition, Fig.~\ref{pic:RQParameters}(c) show that  the increase in the values of quintessence factor ($N_{q}$)  
decrease the temperature of black hole surrounded by quintessence. 
In other words, increasing in the values of quintessence the physical area of this black hole decreases. 
Moreover, in Fig.~\ref{pic:RQParameters}(d), the behaviour of the heat capacity versus  $ r_+ $ is represented.  
Note that, the roots of heat capacity represent a physical limitation points and the 
divergences of the heat capacity demonstrate phase transition critical points of black hole \cite{EslamPanah:2018ums}.
From Fig.~\ref{pic:RQParameters}(d), we see that  the heat capacity of
this black hole will be zero at $ r_{1} $ and $ r_{2} $, which means that it
has two physical limitation points. Furthermore, it diverges at
$ r_+=r_{\infty} =0.825$ and thus it has one phase transition critical point.  
This results that the heat capacity is negative at $ r_+<r_{1} $, which means 
that the black hole system is unstable. 
Then, at $r_{1}<r_+<r_{\infty} $, the heat capacity is positive, which means that it is in
stable phase. Next, at $ r_{\infty}<r_+<r_{2} $, it crashes in
to a negative region (unstable phase) and, at $ r_+>r_{2} $, it
becomes stable.
Further, for the case $N_{q}=0$, the physical limitation points get changed, and there will be only one physical limitation points.
 Moreover, the behaviour of $ \varphi $ as a quantity conjugate to electric charge $ Q $ and 
$ \Psi $ as a quantity conjugate to quintessence parameter $ N_{q} $ versus  $ r_+ $ are shown in Fig.~\ref{pic:RQParameters}(e), (f). 
As can be seen, $ \varphi $ has a maximum value at $ r_{+}= 0.333$, afterword it starts decreasing with higher $ r_+$, 
and $ \Psi $ starts from zero and then it falls into a negative region.
\clearpage
\subsection{Thermodynamic geometry}
\begin{figure}[h]
	\centering
	\subfigure[ $ N_{q}=0.05 $]{
    	\includegraphics[width=0.4\textwidth]{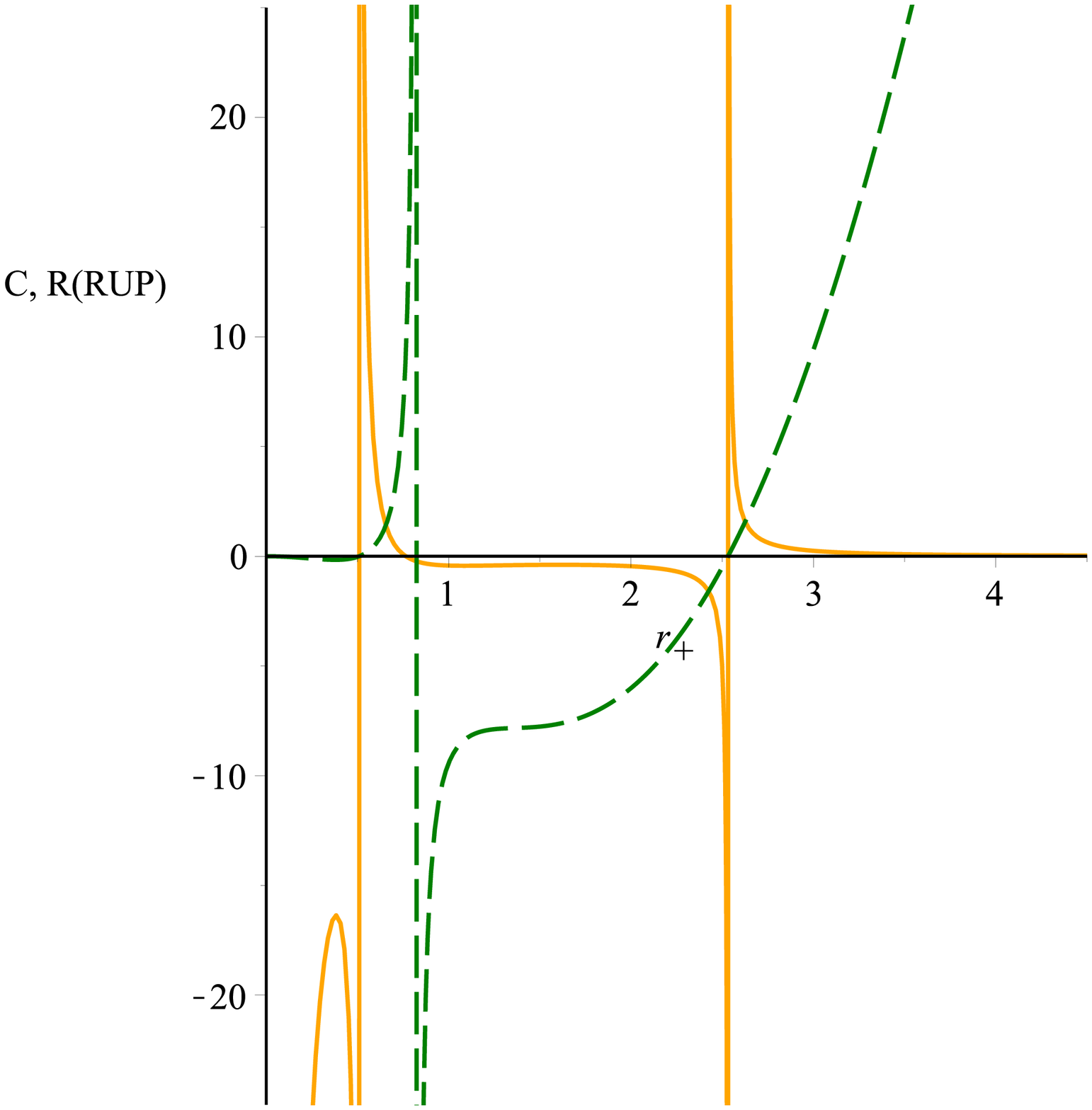}
	}
	 \subfigure[]{
		\includegraphics[width=0.4\textwidth]{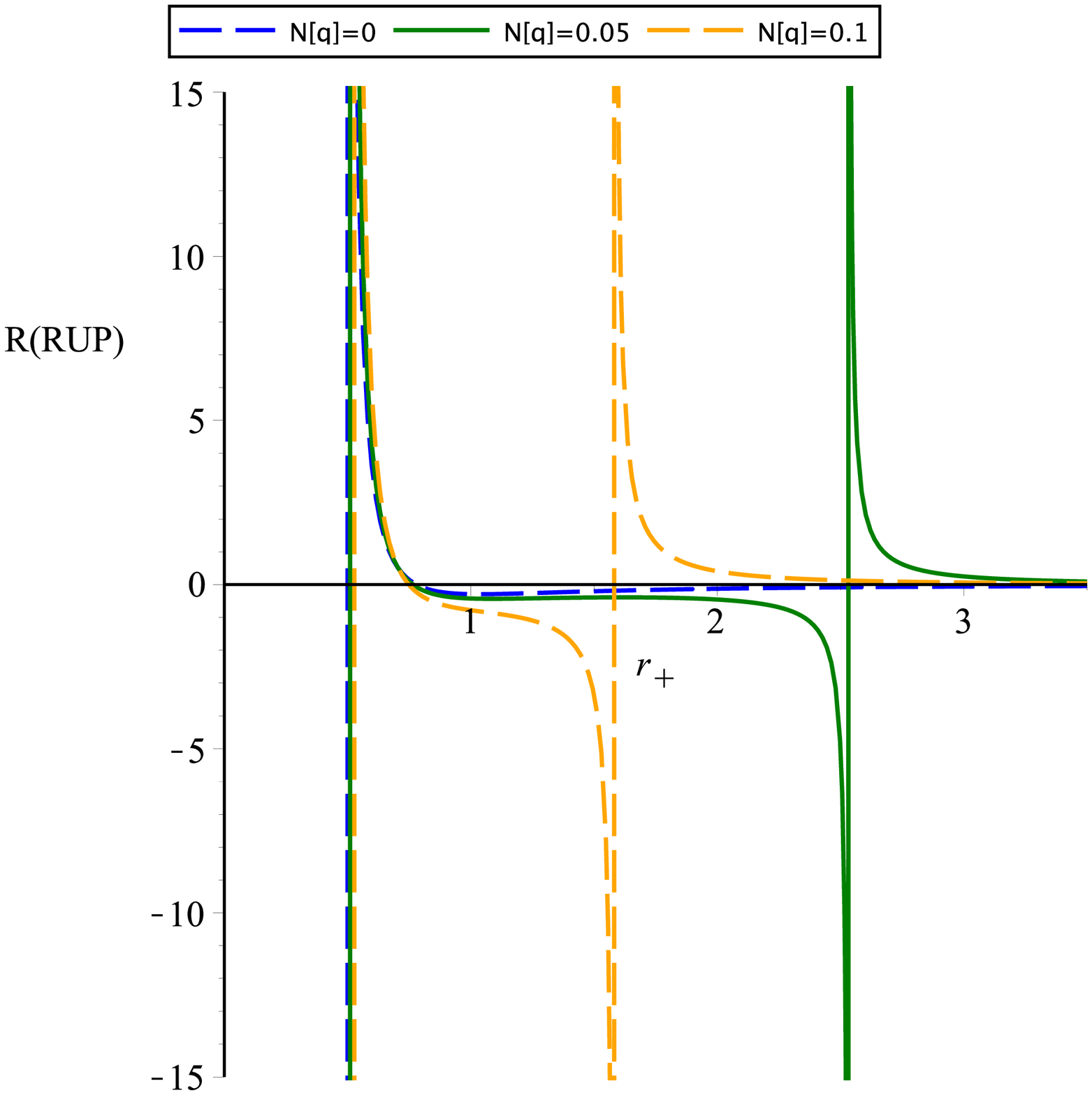}
	}
	\caption{(a) Curvature scalar variation of Ruppeiner metric (orange line) and 
the heat capacity (green dash line) of a black hole surrounded by quintessence field and (b) Curvature scalar variation of Ruppeiner metric for different values of quintessence factor, in terms of horizon radius $ r_{+} $ for $Q=\sqrt{0.25} $. 
}
 \label{pic:RQRup}
\end{figure}

In this section, based on geometric formalism suggested by Weinhold, Ruppiner, Quevedo and HPEM, 
we create the geometric structure 
for a black hole surrounded by quintessence. 
The Weinhold geometry is specified in
 mass representation as~\cite{Weinhold}
 \begin{equation}\label{Weinhold}
 g^{W}_{i j}=\partial _{i}\partial _{j}M(S,N_{q}, Q).
 \end{equation}
In this case, the line element for a black hole surrounded by quintessence is 
\begin{eqnarray}
ds^{2}_{W}&=&M_{SS}dS^{2}+M_{N_{q}N_{q}}dN_{q}^{2}+M_{QQ}dQ^{2}+ 2M_{SN_{q}}dSdN_{q}\nonumber\\
  &+& 2M_{SQ}dSdQ +2M_{N_{q}Q}dN_{q} dQ ,
\end{eqnarray}
 so the matrix
\begin{equation}
g^{W}=\begin{bmatrix}
M_{SS} & M_{SN_{q}} & M_{SQ}\\
M_{N_{q} S} & 0 &0\\
M_{Q S} & 0& M_{Q Q} 
\end{bmatrix}.
\end{equation}
We use the above equations to obtain the curvature scalar 
of the Weinhold metric ($ R^{W} $)
\begin{equation}
R^{W}=0.
\end{equation}
The curvature scalar in the Weinhold formalism is equal to zero, 
thus we cannot describe the phase transition of this thermodynamic system. 
Next, we consider the Ruppiner geometry. Through the conformal property, 
the Ruppiner metric in the thermodynamic system is defined as~\cite{Ruppeiner,Salamon,Mrugala:1984}
\begin{equation}
ds^{2}_{R}=\frac{1}{T}ds^{2}_{W},
\end{equation}
and the relevant matrix is 
\begin{equation}
g^{R}=\left(\frac{4\pi^{\frac{3}{2}}S^{\frac{3}{2}}}{S\pi-\pi^{2}Q^{2}-3S^{2}N_{q}}\right)\begin{bmatrix}
M_{SS} & M_{SN_{q}} & M_{SQ}\\
M_{N_{q} S} & 0 &0\\
M_{Q S} & 0& M_{Q Q} 
\end{bmatrix}.
\end{equation}
Therefore, the curvature scalar of the Ruppiner geometry is given by
\begin{equation}
R^{Rup}=-1/4\,{\frac {\sqrt {S} \left( 17\,{\pi }^{2}{Q}^{2}-9\,{S}^{2}N_{q}-7\,
\pi \,S \right) }{ \left( {\pi }^{2}{Q}^{2}+3\,{S}^{2}N_{q}-\pi \,S
 \right) {\pi }^{3/2}} \left( {\frac {S}{\pi }} \right) ^{-3/2}}.
\end{equation}
The resulting curvature scalar is plotted versus horizon radius 
to investigate thermodynamic phase transition (see Fig.~\ref{pic:RQRup}).
It can be observed from this figure that  the singular points of the curvature scalar 
of Ruppeiner metric coincide   with zero points of the heat capacity.
Moreover, variations of Ruppeiner metric and the heat capacity in terms of different values of 
quintessence factor ($N_{q}$), are shown in Figs.~\ref{pic:RQRup}(b) and \ref{pic:RQParameters}(d).
As one can see from these Figures,
the change in the values of quintessence factor causes the change in zero points of the heat capacity 
and also, singular points of the curvature scalar of Ruppeiner metric. 
But in any case, for both low and high value of the quintessence factor, again the singular points of the curvature scalar 
of Ruppeiner metric entirely coincide  with zero points of the heat capacity.

\begin{figure}[h]
	\centering
	 \subfigure[]{
		\includegraphics[width=0.4\textwidth]{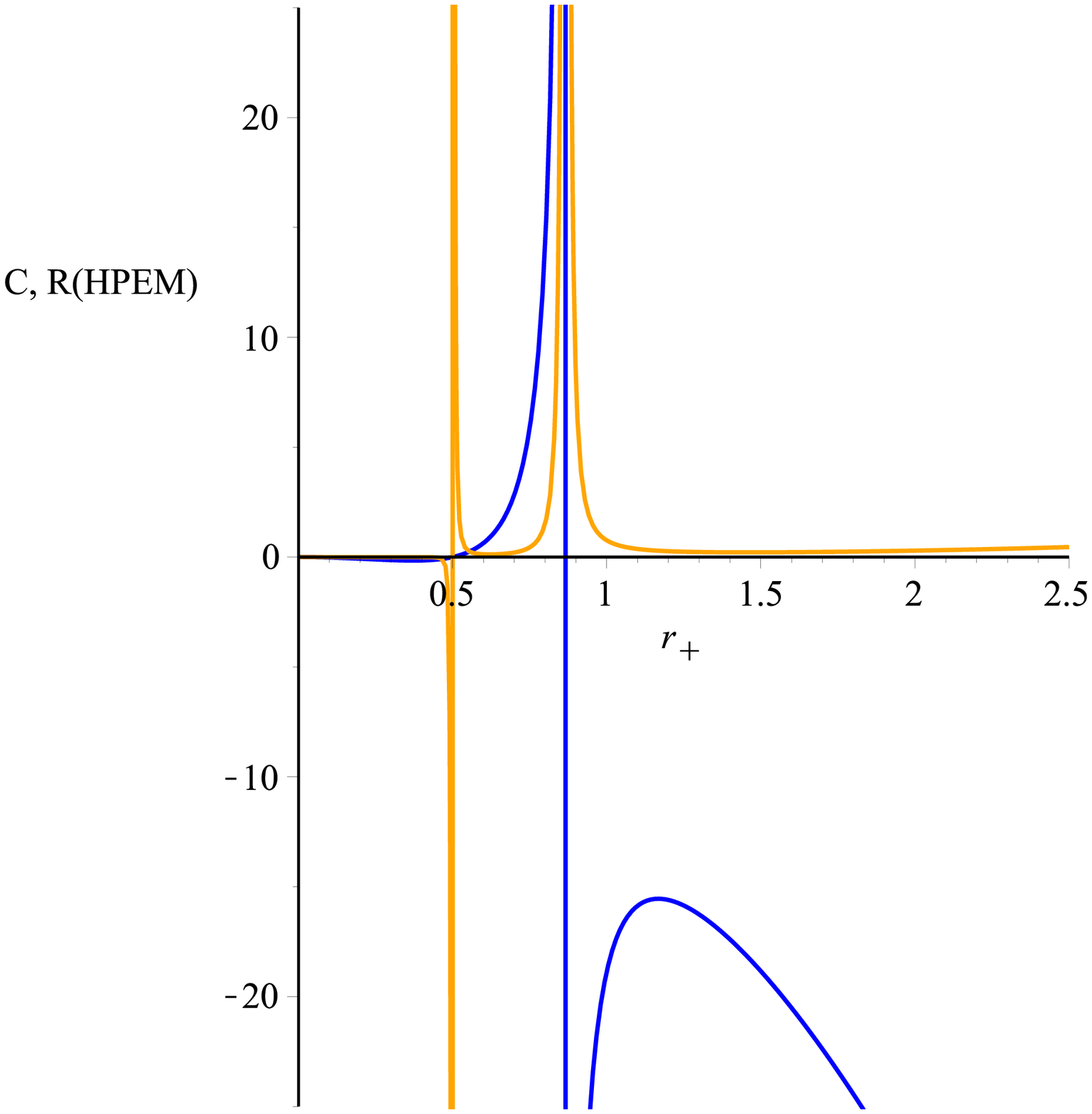}
	}
     \subfigure[]{
		\includegraphics[width=0.4\textwidth]{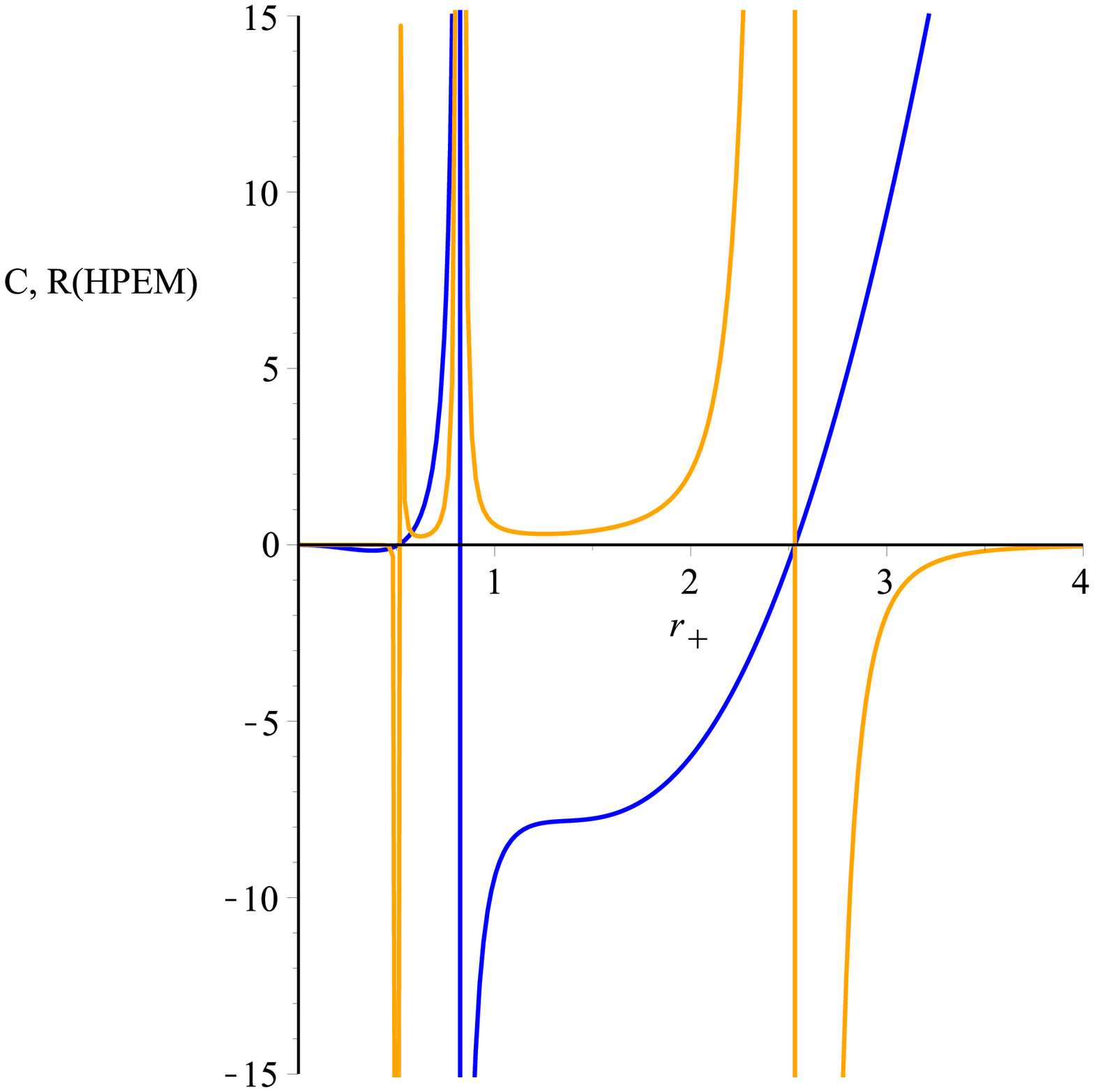}
	}
	\caption{Curvature scalar variation of HPEM metric (orange 
	line) and heat capacity (blue line) of a black hole surrounded by quintessence field 
 in terms of horizon radius $ r_{+} $ for $ Q=\sqrt{0.25} $ and $ N_{q}=0 $, $ N_{q}=0.05 $ for (a) and (b) respectively.}
 \label{pic:RQHPEM}
\end{figure}

Now, we will use the Quevedo and HPEM formalism to discuss the thermodynamic properties of the black hole.
The general form of the metric in Quevedo  formalism is  given as ~\cite{Quevedo,Quevedo:2006xk}:
\begin{equation}\label{GTD}
g=\left(E^{c}\frac{\partial\Phi}{\partial E^{c}}\right)\left(\eta_{ab}\delta^{bc}\frac{\partial^{2}\Phi}{\partial E^{c}\partial E^{d}}dE^{a}dE^{d}\right),
\end{equation}
in which  
\begin{equation}
\frac{\partial\Phi}{\partial E^{c}}=\delta_{cb}I^{b},
\end{equation}
where $ E^{a} $ and $ I^{b} $  are the extensive and intensive 
thermodynamic variables, respectively, and $ \Phi $ is the thermodynamic potential.
Also, the generalized HPEM metric with $ n $ extensive variables ($ n\ge 2 $) has the following form \cite{Hendi:2015rja,Hendi:2015xya,EslamPanah:2018ums}
\begin{equation}
ds_{HPEM}^{2}= \dfrac{SM_{S}}{(\prod_{i=2}^n \frac{\partial^{2} M}{\partial \chi_{i}^{2}})^{3}}\left(-M_{SS}dS^{2}+\sum_{i=2}^n (\frac{\partial^{2} M}{\partial \chi_{i}^{2}})d\chi_{i}^{2}\right) ,
\end{equation}
In which, $ \chi_{i}(\chi_{i}\neq S) $, $ M_{S}=\frac{\partial M}{\partial S} $ and $ M_{SS} =\frac{\partial^{2} M}{\partial S^{2}} $  are extensive parameters.
Moreover, The Quevedo and  HPEM metrics can be written as \cite{Hendi:2015rja,Hendi:2015xya,EslamPanah:2018ums}
\begin{equation}
ds^{2}= \begin{cases}
(SM_{S}+N_{q}M_{N_{q}}+QM_{Q})(-M_{SS}dS^{2}+M_{N_{q}N_{q}}dN_{q}^{2}+M_{QQ}dQ^{2})   & $Quevedo Case I  $ \\

SM_{S}(-M_{SS}dS^{2}+M_{N_{q}N_{q}}dN_{q}^{2}+M_{QQ}dQ^{2})   &$Quevedo Case II  $  .  \\

\dfrac{SM_{S}}{(\frac{\partial^{2} M}{\partial N_{q}^{2}}\frac{\partial^{2} M}{\partial Q^{2}})^{3}}\left(-M_{SS}dS^{2}+M_{N_{q}N_{q}}dN_{q}^{2}+M_{QQ}dQ^{2}\right) & $ HPEM $
 \end{cases}
\end{equation}
These metrics have following denominator for their Ricci scalars \cite{Hendi:2015xya,EslamPanah:2018ums}
\begin{equation}
denom(R)= \begin{cases}
2M_{SS}^{2}M_{N_{q}N_{q}}^{2}M_{QQ}^{2}(SM_{S}+N_{q}M_{N_{q}}+QM_{Q})^{3}   & $Quevedo Case I  $ \\

2S^{3}M_{SS}^{2}M_{N_{q}N_{q}}^{2}M_{QQ}^{2}M_{S}^{3}  & $Quevedo Case II  $  .\\

2S^{3}M_{SS}^{2}M_{S}^{3} & $ HPEM  $
 \end{cases}
\end{equation}
 Solving the mentioned equations, It is evident that in case of Quevedo formalism, we can't find any physical data about the system. 
 But, in case of HPEM metric, as shown in Fig.~\ref{pic:RQHPEM}, divergencies of the Ricci scalar and zero points of
the heat capacity will coincide. 
Of course, there is another divergency of HPEM metric which coincides with the singular points of heat capacity. 
Actually the denominator of the Ricci scalar of HPEM metric only contains numerator and denominator of the heat capacity. 
In other words, divergence points of the Ricci scalar of HPEM metric coincide with both types of phase transitions of the heat capacity. 
Therefore, It seams we can extract more information of HPEM metric compared with Ruppeiner metric.

\section{Black hole surrounded by dust field}\label{section5}
In this section, we consider a black hole surrounded by the dust field. 
So, by putting $w_s=w_d=0$ and $k\lambda=\frac{2}{9}$, 
the Eq. (\ref{mainmetric}) can be written as \cite{12}
\begin{eqnarray}
&ds^2=-f_d(r)dt^2+\frac{dr^2}{f_d(r)}+r^2d \Omega ^2,\\
&f_d(r)=1-\frac{2M}{r}+\frac{Q^2}{r^2}-N_dr .
\end{eqnarray}
\subsection{Thermodynamics}
\begin{figure}[h]
	\centering
	 \subfigure[$ N_{q}=0.05 $]{
		\includegraphics[width=0.38\textwidth]{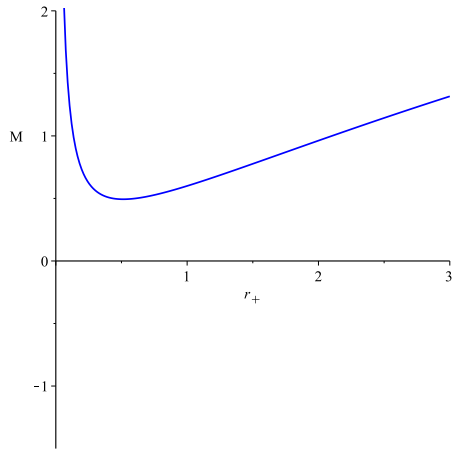}
	}
	\subfigure[]{
		\includegraphics[width=0.38\textwidth]{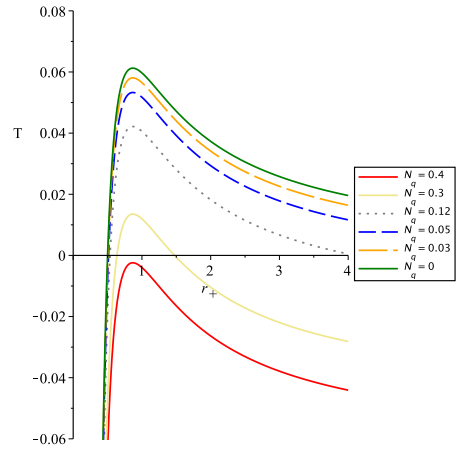}
	}
      \subfigure[]{
		\includegraphics[width=0.38\textwidth]{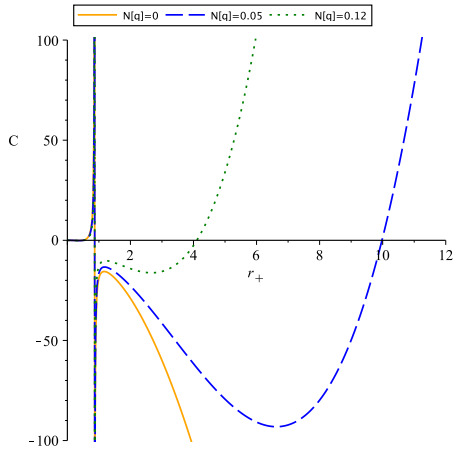}
	}
     \subfigure[$ N_{q}=0.05 $]{
		\includegraphics[width=0.38\textwidth]{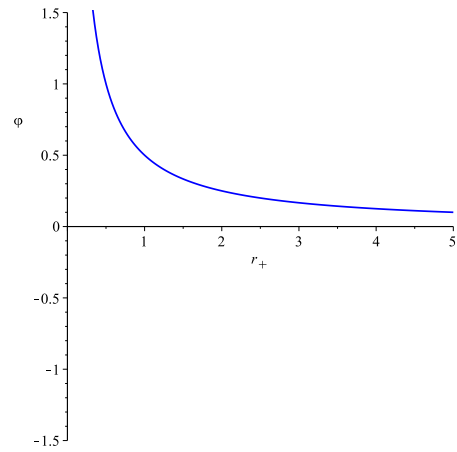}
	}
	\subfigure[$ N_{q}=0.05 $]{
		\includegraphics[width=0.38\textwidth]{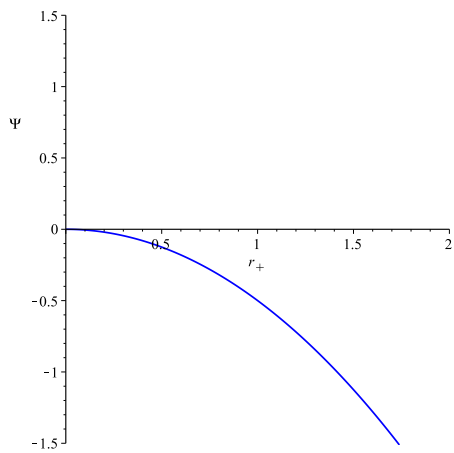}
	}
	\caption{Variations of thermodynamic parameters of a black hole surrounded 
	by dust field in terms of horizon radius $ r_{+} $ for $ Q=\sqrt{0.25} $.}
 \label{pic:RDParameters}
\end{figure}

In this section, we investigate thermodynamical behaviour of 
a black hole surrounded by dust field. In this case, mass of the black hole $M$, in terms of its
entropy $S$, can be written as
\begin{equation}\label{massd}
M(S,N_{d},Q)=\frac{{{\pi ^2}{Q^2} + \pi S - N_{d} {S^\frac{3}{2}}{\pi ^\frac{1}{2}}}}{{2{\pi ^\frac{3}{2}}{S^\frac{1}{2}} }},
\end{equation}
and the first law of thermodynamics for this black hole is
\begin{equation}\label{lowd}
{\it dM}={\it TdS}+\Psi \,dN_{q}+\varphi {\it dQ} ,
\end{equation}
So, by using the above equations (\ref{massd}, \ref{lowd}), 
the thermodynamic parameters can be obtained as 
\begin{equation}
T=- \frac{{{\pi ^2}{Q^2} - \pi S + 2 N_{d} {S^\frac{3}{2}}{\pi ^\frac{1}{2}}}}{{4{\pi ^\frac{3}{2}}{S^\frac{3}{2}}}},
\end{equation}

\begin{equation}
C= - \frac{{2 S \left( {{\pi ^2}{Q^2} - \pi S + 2 N_{d} {S^\frac{3}{2}}{\pi ^\frac{1}{2}}} \right)}}{{3{\pi ^2}{Q^2} - \pi S}} ,
\end{equation}

\begin{equation}
\varphi=\sqrt {\frac{\pi}{S}}Q  ,
\end{equation}

\begin{equation}
\Psi=-\frac{1}{2}\,\frac{S}{\pi} .
\end{equation}
Plots of these thermodynamic parameters in terms of horizon radius $r_+$, 
are shown in Figs.~\ref{pic:RDParameters}.

Fig. \ref{pic:RDParameters}(a) shows that mass of this black hole
has one minimum point at $r_{1}=0.513$. Also, it can be seen from Fig. \ref{pic:RDParameters}(b) 
that temperature of this system is in the positive region at a particular range of event horizon radius $ r_+ $.
Moreover, similar to quintessence case, the increase in the values of dust factor ($N_{d}$) causes 
decrease in the temperature of a black hole surrounded by dust field. 
In addition, Fig. \ref{pic:RDParameters}(c) shows that 
the heat capacity is in the negative region (unstable phase), 
afterword at $ r_{1} $, it becomes zero and Then for $ r_+>r_{1} $, it locate in positive region (stable phase).
Moreover, it can be observed from Fig. \ref{pic:RDParameters}(c) that 
for the case $N_{q}\neq0$, there are two physical limitation points, and for the case $N_{q}=0$, there will be only one physical limitation point.
Furthermore, in Fig.~\ref{pic:RDParameters}(d), (e) variation of $ \varphi $  and $ \Psi $ versus  $ r_+ $ are demonstrated. 
As can be seen, similar to quintessence case, $ \varphi $ has a maximum value at $ r_{+}= 0.333$, afterword it starts decreasing with higher $ r_+$, 
and $ \Psi $ starts from zero and then it falls into a negative region.

\clearpage
\subsection{Thermodynamic geometry}
\begin{figure}[h]
	\centering
	\subfigure[]{
		\includegraphics[width=0.4\textwidth]{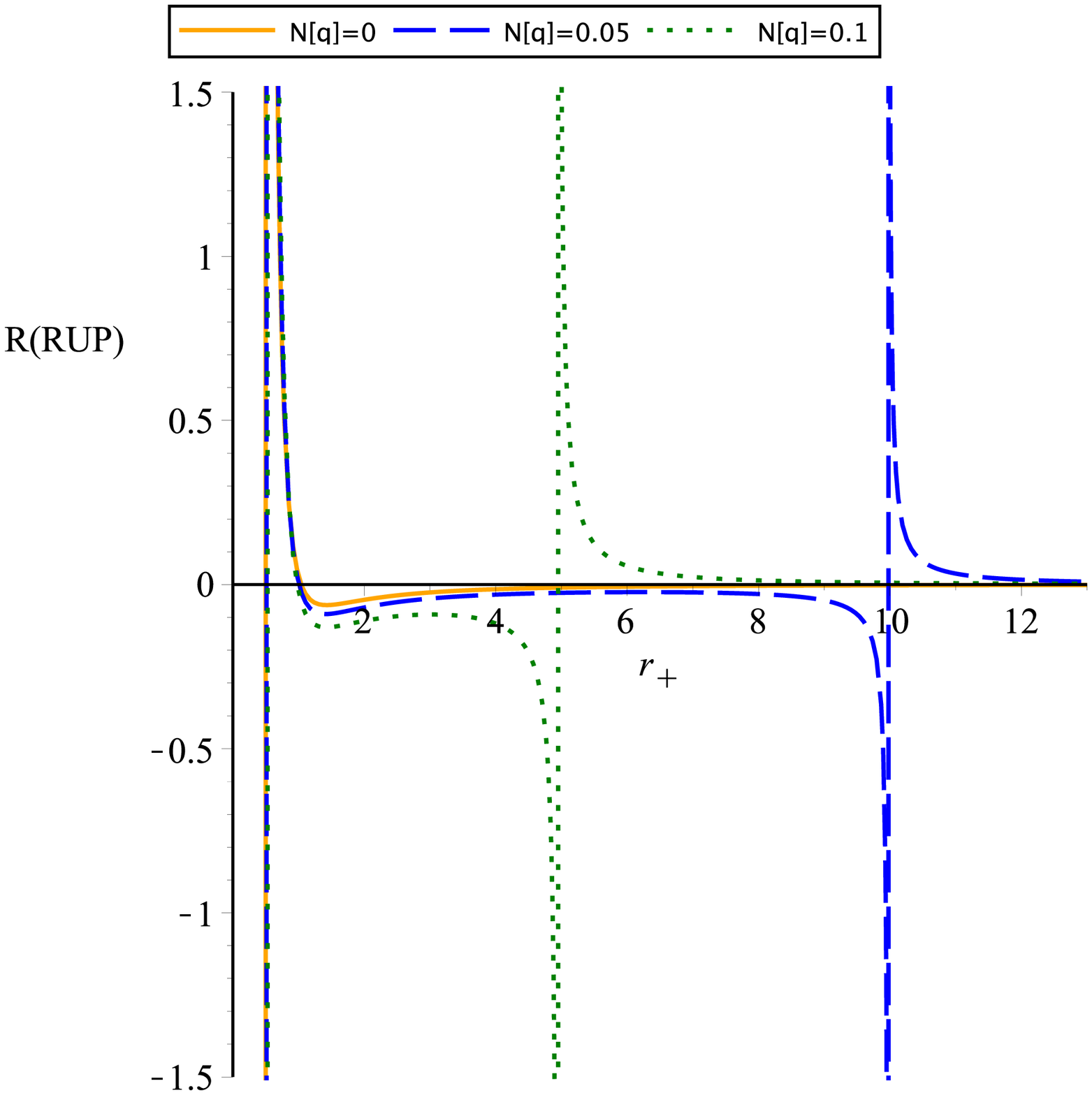}
	}
     \subfigure[]{
		\includegraphics[width=0.4\textwidth]{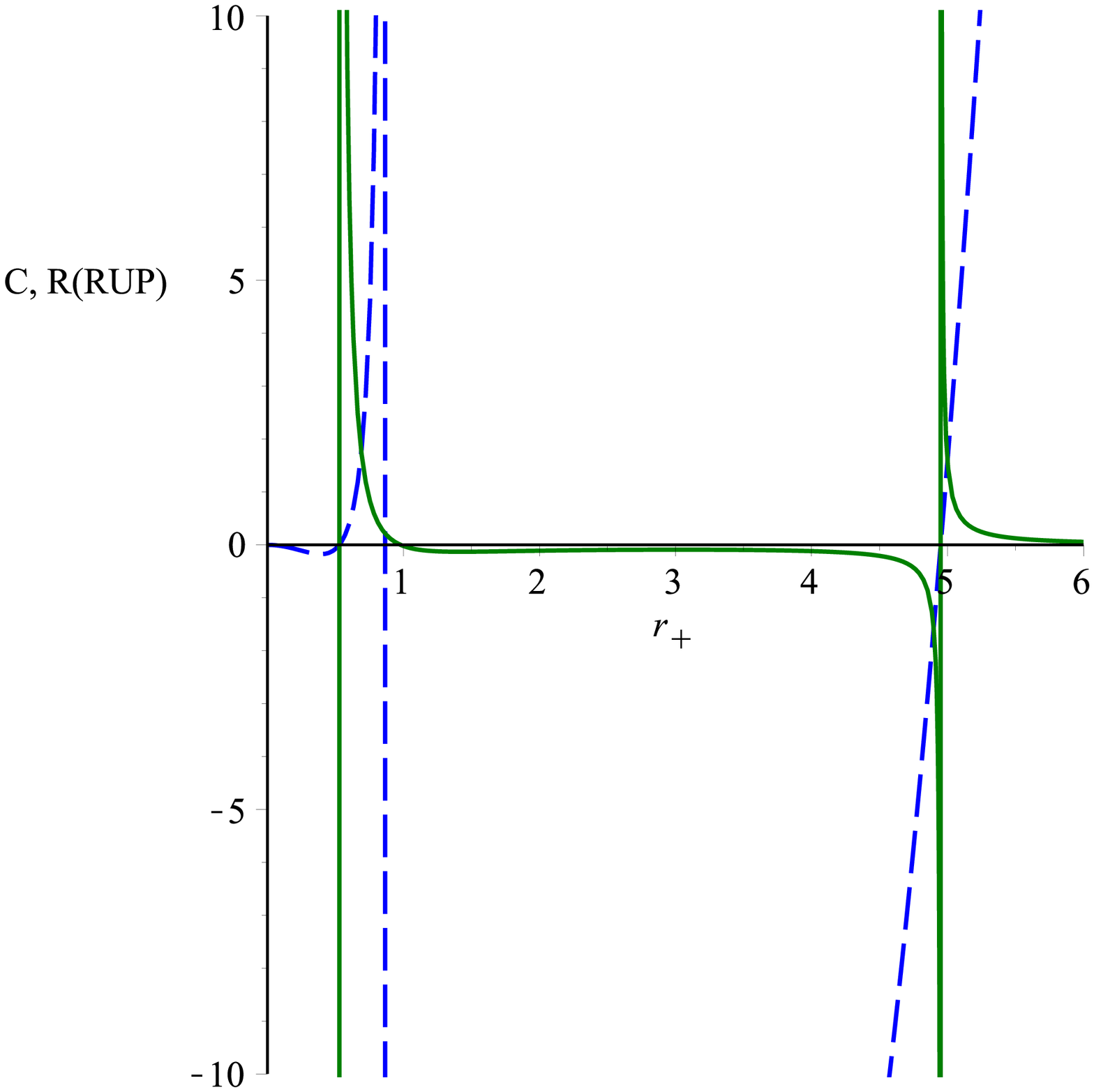}
	}
	\caption{(a) Curvature scalar variation of Ruppeiner metric for different values of dust factor, and (b)
 Curvature scalar variation of Ruppeiner metric (green continuous line)  and heat capacity (blue dash line) of a black hole surrounded by dust field in terms of horizon radius $ r_{+} $ for $ N_{d}=0.1 $, $ Q=\sqrt{0.25} $.}
 \label{pic:RDRup}
\end{figure}

In this section, again we use the geometric formalism of Weinhold, Ruppiner, Quevedo and HPEM 
metrics of the thermal system, and investigate the physical limitation points and phase transition critical
points of a black hole surrounded by dust field. 
we can use the following Weinhold metric ~\cite{Weinhold}:
 \begin{equation}
 g^{W}_{i j}=\partial _{i}\partial _{j}M(S,N_{d}, Q),
 \end{equation}
and write the line element corresponding to 
Weinhold  metric for this system in mass representation as 
\begin{eqnarray}
ds^{2}_{W}&=&M_{SS}dS^{2}+M_{N_{d}N_{d}}dN_{d}^{2}+M_{QQ}dQ^{2}+  2M_{SN_{d}}dSdN_{d}\nonumber\\
 &+&2M_{SQ}dSdQ +2M_{N_{d}Q}dN_{d} dQ .
\end{eqnarray}
Therefore, the relevant matrix is
\begin{equation}
g^{W}=\begin{bmatrix}
M_{SS} & M_{SN_{d}} & M_{SQ}\\
M_{N_{d} S} & 0 &0\\
M_{Q S} & 0& M_{Q Q} 
\end{bmatrix},
\end{equation}
and the curvature scalar of the Weinhold metric for this system is
\begin{equation}
R^{W}=0.
\end{equation}
So,  the Weinhold formalism suggests that  black hole surrounded by dust field is flat and we
cannot investigate the phase transition of this thermodynamic system.
Moreover, we consider Ruppiner formalism in  which line element is conformaly transformed to the Weinhold
metric as \cite{Ruppeiner,Salamon,Mrugala:1984}
\begin{equation}
ds^{2}_{R}=\frac{1}{T}ds^{2}_{W}.
\end{equation}
The matrix corresponding to this metric is 
\begin{equation}
g^{R}=\left(\frac{4\pi^{\frac{3}{2}}S^{\frac{3}{2}}}{S\pi-\pi^{2}Q^{2}-2S^\frac{3}{2}N_{d}\pi^\frac{3}{2}}\right)\begin{bmatrix}
M_{SS} & M_{SN_{d}} & M_{SQ}\\
M_{N_{d} S} & 0 &0\\
M_{Q S} & 0& M_{Q Q} 
\end{bmatrix},
\end{equation}
and the relevant scalar curvature is
\begin{equation}
R^{Rup}=1/4\,{\frac {4\,N_{d}{S}^{3/2}\sqrt {\pi }-13\,{\pi }^{2}{Q}^{2}+3\,\pi \,
S}{ \left( 2\,N_{d}{S}^{3/2}\sqrt {\pi }+{\pi }^{2}{Q}^{2}-\pi \,S
 \right) S}}.
\end{equation}
Plot of this scalar curvature is demonstrated in Fig.~\ref{pic:RDRup}(a).
Moreover, plot of the curvature scalar variation of the Ruppiner metric and heat capacity, in
terms of $ r_+$  is shown in Fig.~\ref{pic:RDRup}(b). 
It can be seen from Fig.~\ref{pic:RDRup}(a), (b) that changes in the value of dust factor cause 
change in the singular points of the curvature scalar of Ruppeiner metric and also in zero points of heat capacity.
Moreover, similar to quintessence case, for both low and high values of dust factor, the singular points of the curvature scalar 
of Ruppeiner metric entirely coincides with zero points of heat capacity.

\clearpage

\begin{figure}[h]
	\centering
	 \subfigure[]{
		\includegraphics[width=0.4\textwidth]{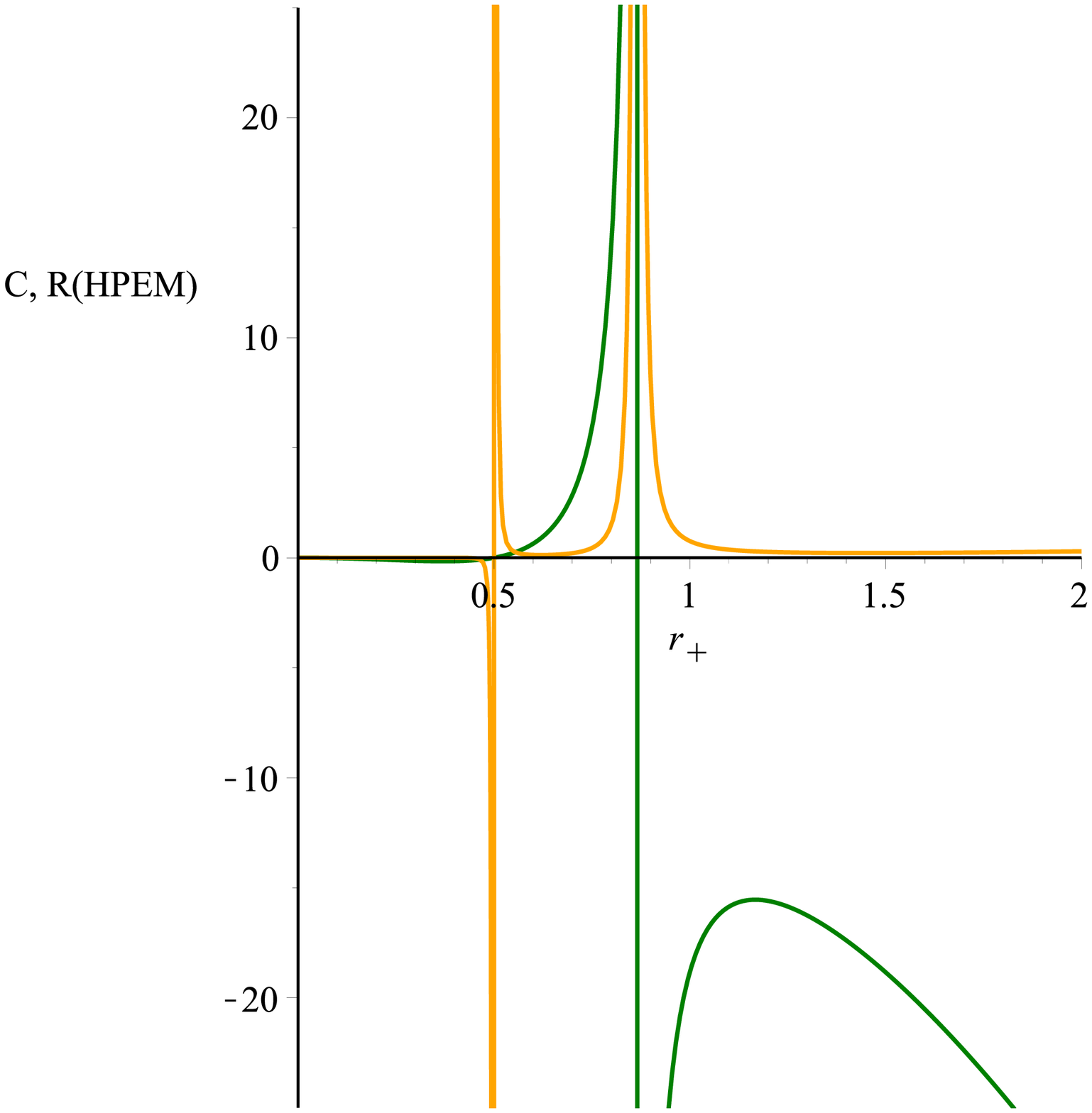}
	}
     \subfigure[]{
		\includegraphics[width=0.4\textwidth]{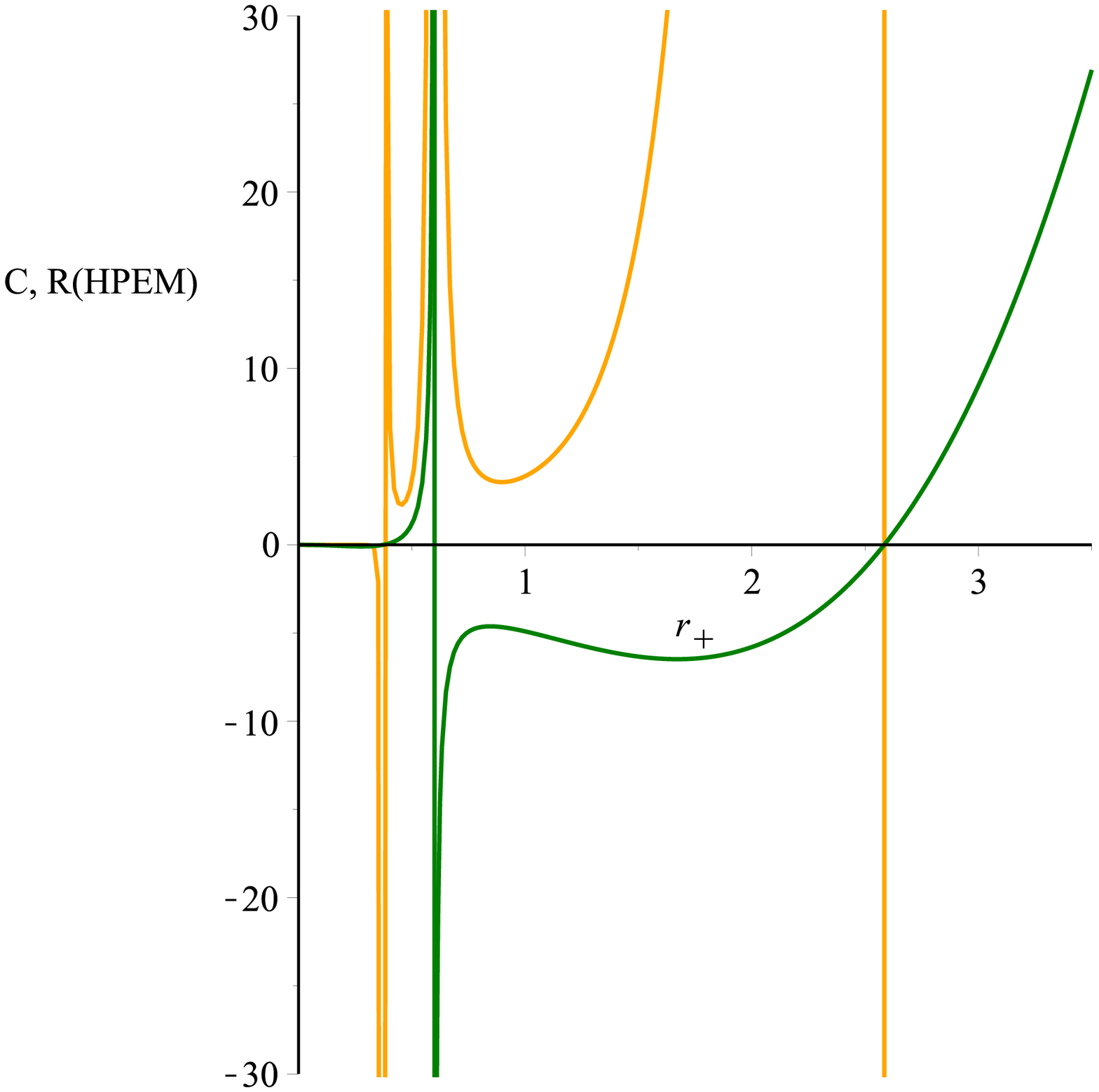}
	}
	\caption{Curvature scalar variation of HPEM metric (orange 
	line) and heat capacity (green line) of a black hole surrounded by dust field 
 in terms of horizon radius $ r_{+} $ for $ Q=\sqrt{0.25} $, $ N_{q}=0 $ and $ Q=\sqrt{0.12} $, $ N_{q}=0.19 $ for (a) and (b) respectively.}
 \label{pic:RDHPEM}
\end{figure}

Now, to extend the analysis, we consider the Quevedo and  HPEM metrics as \cite{Hendi:2015rja,Hendi:2015xya,EslamPanah:2018ums}
\begin{equation}
ds^{2}= \begin{cases}
(SM_{S}+N_{d}M_{N_{d}}+QM_{Q})(-M_{SS}dS^{2}+M_{N_{d}N_{d}}dN_{d}^{2}+M_{QQ}dQ^{2})   & $Quevedo Case I  $ \\

SM_{S}(-M_{SS}dS^{2}+M_{N_{d}N_{d}}dN_{d}^{2}+M_{QQ}dQ^{2})   &$Quevedo Case II  $  .  \\

\dfrac{SM_{S}}{(\frac{\partial^{2} M}{\partial N_{d}^{2}}\frac{\partial^{2} M}{\partial Q^{2}})^{3}}\left(-M_{SS}dS^{2}+M_{N_{d}N_{d}}dN_{d}^{2}+M_{QQ}dQ^{2}\right) & $ HPEM $
 \end{cases}
\end{equation}
These metrics have following denominator for their Ricci scalars \cite{Hendi:2015xya,EslamPanah:2018ums}
\begin{equation}
denom(R)= \begin{cases}
2M_{SS}^{2}M_{N_{d}N_{d}}^{2}M_{QQ}^{2}(SM_{S}+N_{d}M_{N_{d}}+QM_{Q})^{3}   & $Quevedo Case I  $ \\

2S^{3}M_{SS}^{2}M_{N_{d}N_{d}}^{2}M_{QQ}^{2}M_{S}^{3}  & $Quevedo Case II  $  .\\   

2S^{3}M_{SS}^{2}M_{S}^{3} & $ HPEM  $  
 \end{cases}
\end{equation}
 Solving the above equations, again similar to the quintessence section for the case of Quevedo formalism, we can't find any physical data about the system. 
 However, in case of HPEM metric, as shown in Fig.~\ref{pic:RDHPEM}, divergencies of the Ricci scalar coincide with zero points of
the heat capacity. 
But, In comparison with Ruppeiner metric, we can see another divergency of HPEM metric which coincides with the singular points of heat capacity. 
So, divergence points of the Ricci scalar of HPEM metric coincide with both types of phase transitions of the heat capacity.

\section{Concluding remarks}\label{section6}
One possible explanation to the negative pressure in an expanding universe is that the universe is filled with a peculiar kind
of perfect fluid, where the proportion between the pressure and energy density is between
$−1$ and $−1/3$. If this is so, then it becomes important to study the interaction of the strong gravity objects such as black holes  with such fluid. This  idea is proposed by Kiselev \cite{kiselev}, and has further been generalized to the Rastall model of
gravity  \cite{12}.

In this paper, we have considered a black hole surrounded
by a generic perfect fluid in Rastall theory to emphasize its thermodynamics. To
make the analysis clear, we have focused on two  special cases of the  perfect fluid. 
First, we have derived an expression for the mass of   black hole surrounded
by quintessence field. The resulting mass of this system  has a minimum value for small horizon radius $r_+$, afterword it arrives to its maximum value for a bit higher value of $r_+$, then vanishes for larger $r_+$. By exploiting standard thermodynamic  relations,
we have derived the Hawking temperature and heat capacity. It is known that negative heat capacity corresponds to instability, however the positive value of heat capacity 
describe stable black holes. In order to study the behavior  of thermodynamic entities 
obtained for black hole surrounded by both quintessence field and dust field, we have plotted the graphs
with respect to event horizon. For the figure, it is clear that temperature of black hole surrounded
by quintessence field in Rastall theory remains positive only for the
specific values of event horizon for quintessence factor $ N_q = 0.05$.  It is also found that
the increasing in the values of quintessence the physical area of this black hole decrease. 
For heat capacity plot, we have found that there exist  
two physical limitation points for the system at two different values of event horizons. Here we note that the stability do not occur corresponding to all horizon radius. Further, we have studied the
geometric structure of the black hole surrounded by quintessence field. This is done by calculating the Weinhold, Ruppiner, Quevedo and HPEM curvature scalars. We find that Weinhold curvature scalar vanishes and in this situation one can not describe the
phase transition.  As well as, we have derived the curvature scalar of Quevedo metric and haven't found any physical data about the system. The curvature scalar for the Ruppiner and the HPEM geometry does not vanishes. The curvature scalar for the Ruppiner and the HPEM geometry suggests that the singular points coincide with zero points of the heat capacity. Also, we have observed another divergency of HPEM metric which coincides with the singular points of heat capacity. So, divergence points of the Ricci scalar of HPEM metric coincide with both types of phase transitions of the heat capacity. Therefore, It can be extracted more information from HPEM metric compared with other mentioned metrics.

Moreover, the thermodynamic properties of black hole surrounded by dust field in Rastall 
gravity are also studied. Following the earlier case, we have calculated the mass, Hawking temperature and heat capacity here also. To emphasize the behavior of these quantities, we have plotted the graphs with respect to event horizon. From the figure, we have found that  the mass of the system has one minimum point at  particular value of horizon radius. The behavior of  temperature with horizon radius is similar to  the quintessence case, i.e., increasing in the values of quintessence the physical area of this black hole decrease. From the plot, we have seen  that stability of black hole increases with the size of black hole upto a certain point. After that point a phase transition occurs and converts the black hole in to the more unstable state. The geometric properties of black hole is studied in this case also. It would be interesting to discuss the effect of thermal fluctuations on thermodynamics of the black hole surrounded by Rastall gravity.

\end{document}